\documentclass[11pt]{article}

\usepackage[utf8]{inputenc}
\usepackage[T1]{fontenc}
\usepackage{lmodern}
\usepackage{geometry}
\usepackage{graphicx}
\usepackage{booktabs}
\usepackage{array}
\usepackage{xcolor}
\usepackage{listings}
\usepackage{authblk}
\usepackage{natbib}
\usepackage{url}
\usepackage{hyperref}

\definecolor{darkblue}{RGB}{0,0,120}
\hypersetup{colorlinks=true, citecolor=darkblue, linkcolor=darkblue, urlcolor=darkblue}

\title{Snyk VulnBench JS 1.0: Can LLMs Find the Same Bugs Twice?}
\author[1]{Liran Tal}
\author[1]{Johannes Kloos}
\author[1]{Arsenii Rudich}
\author[1]{Stephen Thoemmes}
\author[1]{Manoj Nair}
\affil[1]{Snyk}
\date{June 11, 2026}

\lstdefinelanguage{JavaScript}{
  keywords={const,let,var,function,return,if,else,true,false},
  sensitive=true,
  comment=[l]{//},
  morecomment=[s]{/*}{*/},
  morestring=[b]",
  morestring=[b]',
  morestring=[b]`
}

\begin{document}

\maketitle

\begin{abstract}
We ran 300 repeated vulnerability-finding scans to measure how repeatable agentic large language model (LLM) security review is on the same JavaScript code, prompt, and benchmark harness. The headline result is that LLM security findings were unevenly repeatable: reference-matched findings were stable, but extra model reports varied heavily from run to run. Across 250 model runs, 80 of 161 unique unmatched findings appeared in only one of five identical repetitions, while only 22 appeared in all five. By contrast, when Claude matched a Snyk Code reference finding, the behavior was much more stable: 134 of 158 unique reference-matched findings appeared in all five repetitions. The benchmark also shows complementarity. Models consistently found familiar, high-signal exploit shapes, and in one case surfaced a likely Snyk Code product gap. Snyk Code static application security testing (SAST) was deterministic and better at systematically enumerating repeated data-flow sinks. The results support combining agentic LLM review with deterministic SAST rather than treating either technique as a replacement for the other.
\end{abstract}

\section{Introduction}

Coding agents are now part of the development loop. They write code, modify pull requests, explain changes, and increasingly perform security review before a human reads the diff. That makes reliability a product question: if the same agent sees the same vulnerable code twice, does it report the same security issues twice?

Traditional SAST tools are built to be deterministic. If the code and rules are unchanged, the output should be unchanged. LLMs are different. They can reason about unfamiliar code, describe risk in useful prose, and sometimes spot issues that a static analyzer misses. But they can also vary across runs, over-report adjacent concerns, or stop after finding one representative example of a repeated pattern.

Snyk VulnBench JS 1.0 was designed to quantify that behavior. The benchmark uses small JavaScript and Express applications so every run is inspectable. The point is not to simulate an entire monorepo. The point is to make model behavior measurable under repeated, controlled conditions.

The model configurations were run through a Claude Code harness using the TypeScript Agent SDK \cite{anthropic-claude-code,anthropic-agent-sdk}. Snyk Code provides the deterministic SAST reference set for this release \cite{snyk-code-docs}. The benchmark should therefore be read as a repeatability and agreement study, not as a universal vulnerability-detection leaderboard.

\paragraph{Key highlights.}
\begin{itemize}
  \item The highest-recall LLM configuration found only 81\% of Snyk Code reference vulnerabilities.
  \item The best-scoring LLM configuration reached 75.4\% Snyk-reference F1, leaving a 24.6-point gap against deterministic SAST reference reproduction.
  \item Nearly 50\% of LLM-only vulnerability reports appeared in just one of five identical scans.
  \item The highest-recall LLM also had the noisiest queue: 41\% of its reports fell outside the Snyk Code reference set.
  \item In the largest app-like fixture, the best model scored only 40.0\% Snyk-reference F1 and repeatedly missed path traversal and resource-limit vulnerabilities.
\end{itemize}

\section{Methodology}

\subsection{Benchmark Design}

The benchmark contains 10 JavaScript fixture projects with 44 Snyk Code reference findings. Each fixture is a small Express-based application, ranging from compact single-file snippets to a larger todo app with server routes, database state, uploads, and frontend JavaScript.

We evaluated six configurations, shown in Table~\ref{tab:configs}. The Claude model identifiers and model family behavior are described by Anthropic's model documentation \cite{anthropic-models-overview}.

\begin{table}[htbp]
\centering
\caption{Benchmark configurations.}
\label{tab:configs}
\begin{tabular}{llr}
\toprule
Configuration & Type & Repetitions per task \\
\midrule
Snyk Code SAST & Command baseline & 5 \\
Claude Opus 4.6 Medium & Model via Claude Code harness & 5 \\
Claude Opus 4.6 High & Model via Claude Code harness & 5 \\
Claude Opus 4.7 Max & Model via Claude Code harness & 5 \\
Claude Sonnet 4.6 Medium & Model via Claude Code harness & 5 \\
Claude Sonnet 4.6 High & Model via Claude Code harness & 5 \\
\bottomrule
\end{tabular}
\end{table}

Each configuration ran each task five times: 10 tasks, 6 configurations, and 5 repetitions, for 300 total runs. The model configurations used the same direct audit prompt and returned findings as structured JSON. The model could read the project files, but not the \texttt{findings.json} reference file.

Snyk Code defines the reference set for this benchmark. That means its 100\% score is not an accuracy claim about all possible vulnerabilities in the projects. It means Snyk Code reproduced its own reference findings deterministically across repeated runs. We use that reference set to measure model agreement, model variance, and where model behavior diverges.

\subsection{Scoring}

The scorer is intentionally lenient: a model finding is credited if it reports the same vulnerability type as a reference finding. It does not need to match the same file, line, severity, or source-to-sink path. We report Snyk-reference F1: the harmonic mean of precision and recall when Snyk Code findings are treated as the reference set. It is useful as an agreement metric, but it is not the main story.

Because this benchmark does not use an independent, exhaustively adjudicated ground truth set, Snyk-reference F1 should not be read as true vulnerability-detection accuracy. A model with lower agreement against the Snyk Code reference set could still score better against an independent ground truth if some of its unmatched findings are valid, or if it avoids issues that the reference set overstates. In this report, Snyk-reference F1 answers a narrower question: how closely and how repeatably do model findings align with the Snyk Code reference findings?

\section{Results}

\subsection{LLM Repeatability Varied by Model Configuration}

At the configuration level, repeatability shows up as the relationship between score and variance. The stronger outcome is toward the upper-left: high agreement with the reference set and low repeated-run variance. Snyk Code SAST sits at that corner with 100.0\% Snyk-reference F1 and 0.0 percentage-point standard deviation because it reproduced its reference set deterministically. The Claude model configurations spread downward and to the right, with Claude Sonnet 4.6 High showing the largest headline variance at 3.5 percentage points.

\begin{figure}[htbp]
\centering
\includegraphics[width=0.95\linewidth]{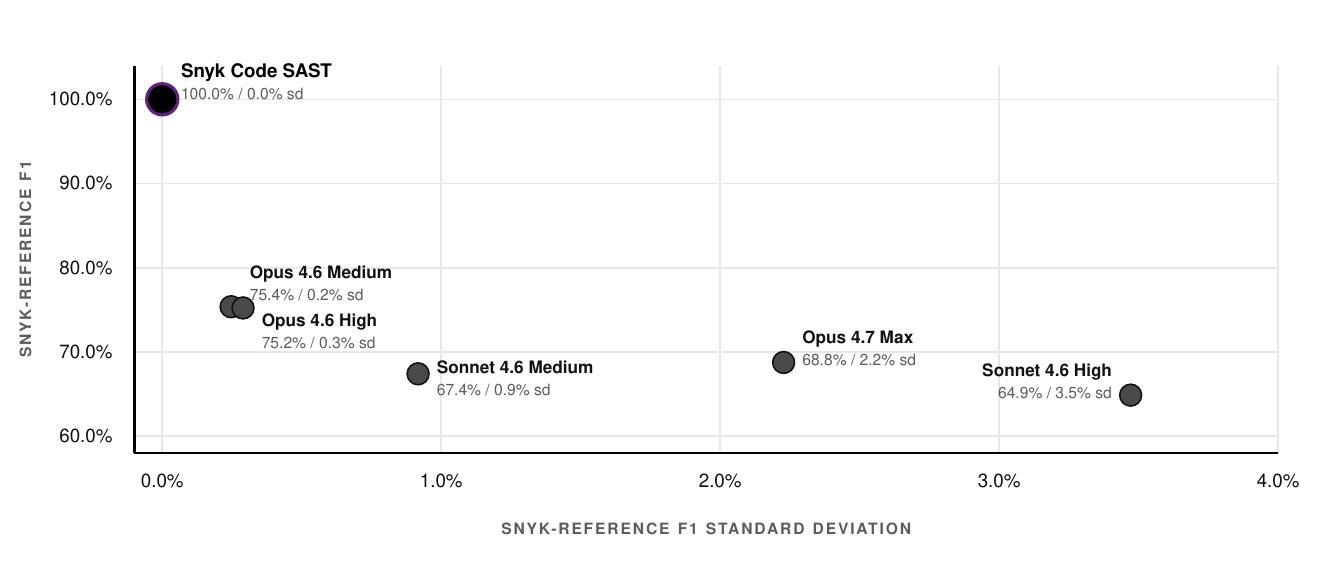}
\caption{Snyk-reference F1 plotted against headline Snyk-reference F1 standard deviation. Better points move toward the top-left: higher agreement score with lower repeated-run variance. Snyk Code SAST is highlighted in purple at zero variance.}
\label{fig:score-stability}
\end{figure}

The scatter makes two things visible at once. First, Snyk Code's role in this benchmark is deterministic reference reproduction, not probabilistic review. Second, the model configs do not line up by cost or recency: Claude Opus 4.6 Medium and High sit close together with low variance and the strongest model Snyk-reference F1, while Claude Opus 4.7 Max and Claude Sonnet 4.6 High are farther right, meaning their repeated runs moved more.

\begin{quote}
\textbf{Highlight:} The best-scoring LLM configuration reached 75.4\% Snyk-reference F1, leaving a 24.6-point gap against deterministic SAST reference reproduction.
\end{quote}

Across all model configurations, 80 of 161 unique unmatched finding signatures appeared in only one of five repeated runs. That aggregate explains part of the variance story, but the more useful view is model-by-model.

\begin{figure}[htbp]
\centering
\includegraphics[width=0.95\linewidth]{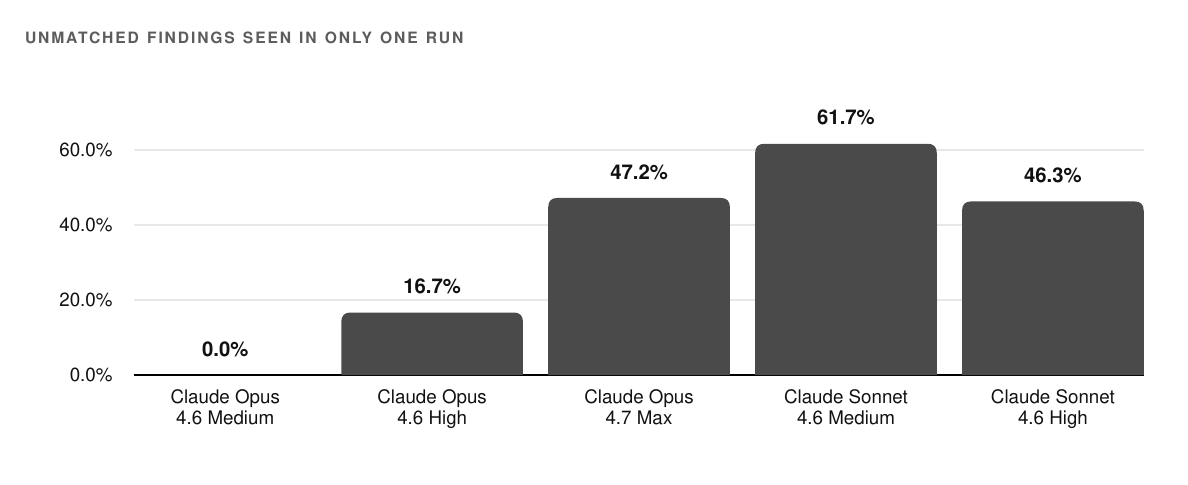}
\caption{Share of each model configuration's unique unmatched finding signatures that appeared in only one of five repeated runs. Signature = task + vulnerability type + file + line, grouped by model config.}
\label{fig:one-run-unmatched}
\end{figure}

Table~\ref{tab:stability-by-model} provides exact values for the preceding chart, plus the two stability counters that matter most.

\begin{table}[htbp]
\centering
\caption{Unmatched and reference-matched finding stability by model configuration.}
\label{tab:stability-by-model}
\small
\resizebox{\textwidth}{!}{%
\begin{tabular}{lrrrr}
\toprule
Model configuration & Unique unmatched & Seen in 1 of 5 & Seen in all 5 & Reference-matched in all 5 \\
\midrule
Claude Opus 4.6 Medium & 5 & 0.0\% & 60.0\% & 100.0\% \\
Claude Opus 4.6 High & 6 & 16.7\% & 50.0\% & 96.2\% \\
Claude Opus 4.7 Max & 36 & 47.2\% & 16.7\% & 74.3\% \\
Claude Sonnet 4.6 Medium & 60 & 61.7\% & 8.3\% & 80.6\% \\
Claude Sonnet 4.6 High & 54 & 46.3\% & 9.3\% & 80.6\% \\
\bottomrule
\end{tabular}
}
\end{table}

The instability is not evenly distributed across models. Claude Sonnet 4.6 Medium produced the largest unmatched finding surface, with 60 unique unmatched signatures; 37 of those appeared in only one of five runs. Claude Sonnet 4.6 High and Claude Opus 4.7 Max showed a similar pattern: many extra reports appeared once and did not recur. In contrast, Claude Opus 4.6 Medium produced few one-off findings, with all of its extra reports found in two or more runs.

\begin{quote}
\textbf{Highlight:} Claude Sonnet 4.6 Medium produced the most one-off extra vulnerability reports: 61.7\% of its LLM-only reports appeared in just one of five runs.
\end{quote}

The Opus 4.6 configurations behaved differently. They produced far fewer unmatched findings, and their extra reports were more stable. That does not make every extra report correct, but it changes the operational interpretation: fewer surprise findings, fewer one-off claims, and less triage churn.

\begin{figure}[htbp]
\centering
\includegraphics[width=0.95\linewidth]{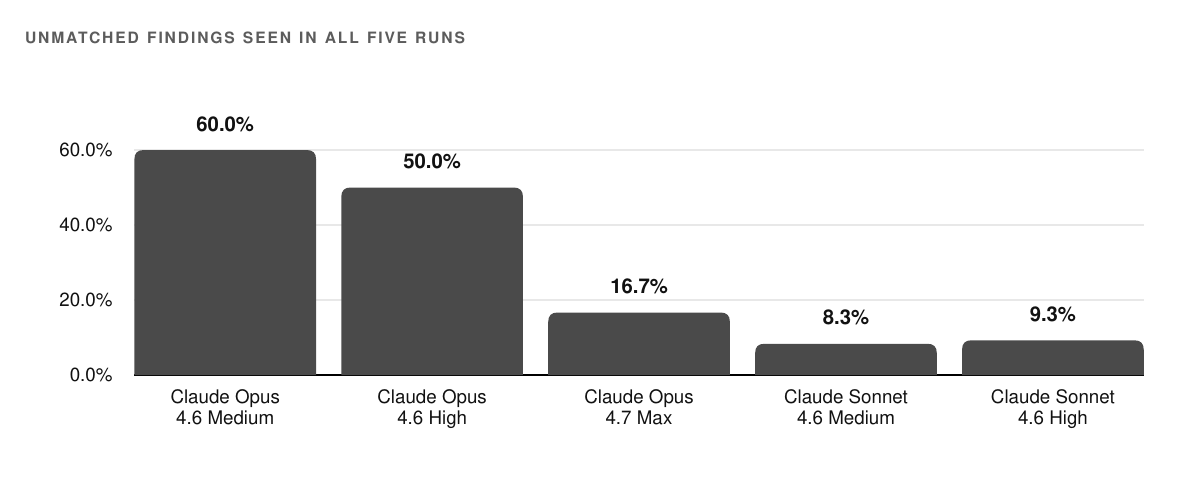}
\caption{Share of unique unmatched finding signatures that appeared in all five repeated runs for each model configuration.}
\label{fig:stable-unmatched}
\end{figure}

The matched side tells a different story. When a model found a Snyk Code reference finding, it usually found it repeatedly. Claude Opus 4.6 Medium matched 25 unique reference findings and repeated all 25 across five runs. Claude Opus 4.6 High repeated 25 of 26. Even the noisier Sonnet configurations repeated 29 of 36 reference-matched findings.

\begin{figure}[htbp]
\centering
\includegraphics[width=0.95\linewidth]{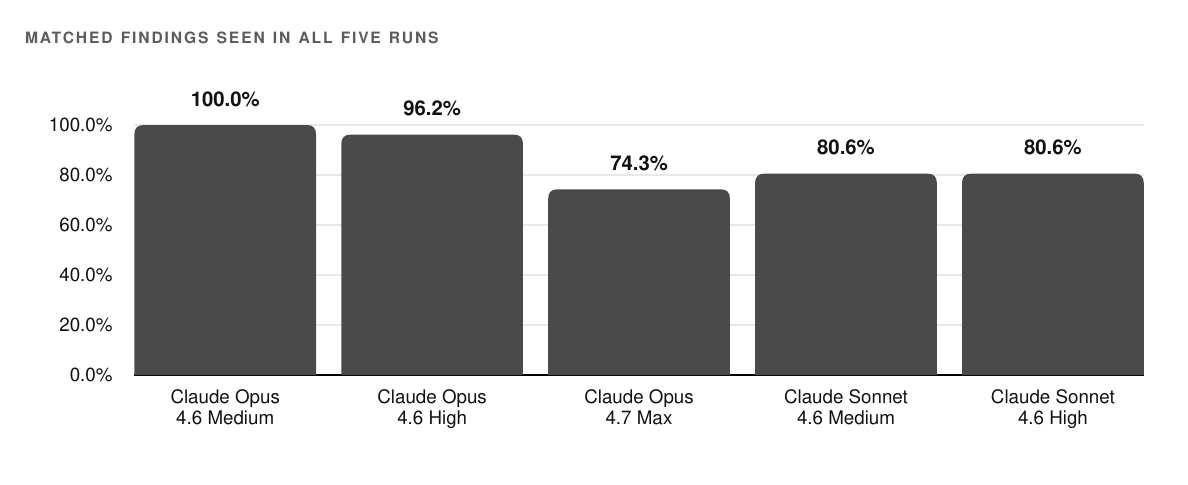}
\caption{Share of unique Snyk Code reference findings matched in all five repeated runs for each model configuration.}
\label{fig:stable-matched}
\end{figure}

\begin{figure}[htbp]
\centering
\includegraphics[width=0.95\linewidth]{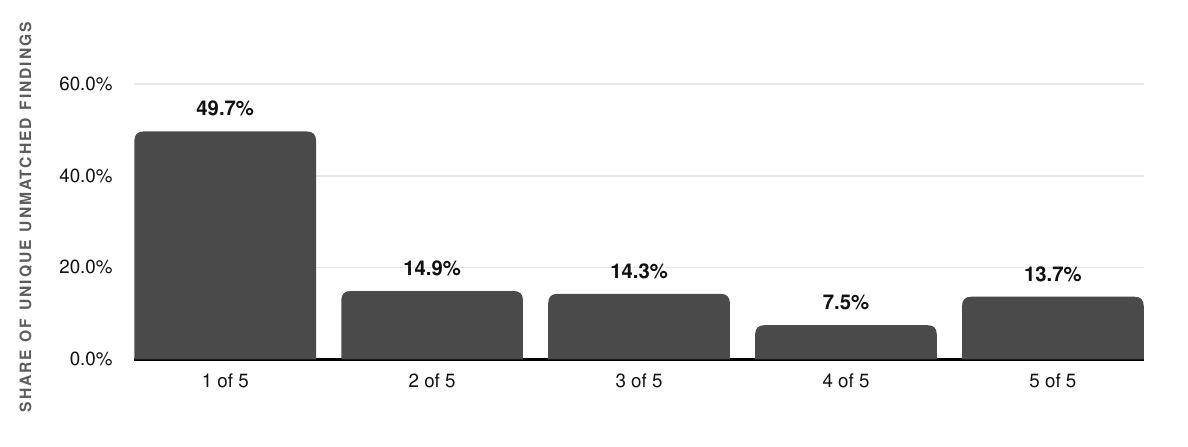}
\caption{Distribution of unique unmatched model findings by how often the same finding signature appeared across five repetitions of the same task and model config. Signature = task + config + vulnerability type + file + line.}
\label{fig:unmatched-repeatability}
\end{figure}

Across all model configurations, 134 of 158 unique reference-matched findings appeared in all five repetitions, or 84.8\%. That means when a model spotted a known Snyk Code reference finding, it usually did so reliably. The contrast with Figure~\ref{fig:unmatched-repeatability} is the main result: true positives were usually stable, while extra non-reference reports were much noisier.

\begin{quote}
\textbf{Highlight:} Of the Snyk Code reference vulnerabilities LLMs found, 85\% were reported consistently across all five identical scans.
\end{quote}

Nearly half of unique unmatched model findings appeared in only one of five identical repetitions. That is a practical reliability problem: a developer could get a materially different review queue depending on which run happened to execute.

\begin{quote}
\textbf{Highlight:} Only 14\% of extra LLM vulnerability reports appeared in every run, making the non-reference review queue much less repeatable.
\end{quote}

\begin{table}[htbp]
\centering
\caption{Distribution of unique unmatched findings across repeated runs.}
\label{tab:unmatched-distribution}
\begin{tabular}{lrr}
\toprule
Repetition frequency & Unique unmatched findings & Share \\
\midrule
1 of 5 runs & 80 & 49.7\% \\
2 of 5 runs & 24 & 14.9\% \\
3 of 5 runs & 23 & 14.3\% \\
4 of 5 runs & 12 & 7.5\% \\
5 of 5 runs & 22 & 13.7\% \\
\bottomrule
\end{tabular}
\end{table}

\subsection{LLM Agents and SAST Found Different Security Gaps}

The most useful interpretation is not ``LLM versus SAST.'' It is ``LLM plus SAST catches different failure modes.''

\begin{figure}[htbp]
\centering
\includegraphics[width=1.0\linewidth]{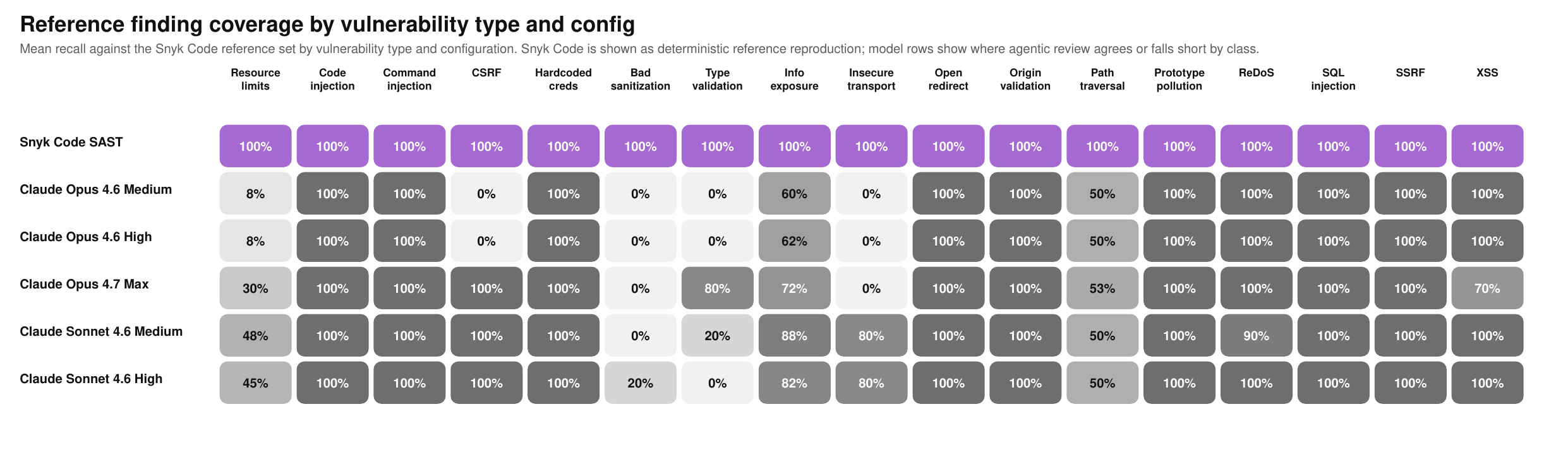}
\caption{Mean recall against the Snyk Code reference set by vulnerability type and configuration. Snyk Code is shown as deterministic reference reproduction; model rows show where agentic review agrees or falls short by class.}
\label{fig:reference-coverage}
\end{figure}

The clearest way to see this is by vulnerability class. Figure~\ref{fig:reference-coverage} shows mean recall against the Snyk Code reference set by vulnerability type and configuration. Snyk Code appears as 100\% because it defines and reproduces the reference set. The model rows show where agentic review agrees with that reference set and where it falls short.

The model configurations were strongest on familiar, high-signal exploit shapes: command injection, code injection, hardcoded credentials, SQL injection, server-side request forgery (SSRF), open redirect, prototype pollution, and regular-expression denial of service (ReDoS) were often found cleanly. They were weaker on resource-limit findings, improper sanitization, type validation, insecure transport, framework information exposure, and repeated path traversal flows.

\begin{quote}
\textbf{Highlight:} LLMs were weaker on systematic SAST classes: resource-limit findings, framework information exposure, insecure transport, sanitization and type-validation issues, and repeated path traversal flows.
\end{quote}

That pattern is visible in \texttt{js-project-tigerteam}, where every model configuration consistently found the hardcoded database password, reflected cross-site scripting (XSS), path traversal, and command injection across all 25 model repetitions.

\begin{lstlisting}[language=JavaScript]
app.get("/greet", (req, res) => {
  const name = req.query.name;
  res.send(`<html><body><h1>Hello, ${name}!</h1></body></html>`);
});

app.get("/file", (req, res) => {
  const filename = req.query.filename;
  const basePath = "/var/app/public/";
  fs.readFile(basePath + filename, "utf8", (err, data) => {
    if (err) return res.status(404).send("Not found");
    res.send(data);
  });
});

app.get("/ping", (req, res) => {
  const host = req.query.host;
  exec("ping -c 1 " + host, (err, stdout, stderr) => {
    if (err) return res.status(500).send("Error");
    res.send(`<pre>${stdout}</pre>`);
  });
});
\end{lstlisting}

But the same fixture included a SQL-shaped mock helper:

\begin{lstlisting}[language=JavaScript]
function dbQuery(sql) {
  console.log("Query:", sql);
  return [];
}

app.get("/users", (req, res) => {
  const username = req.query.username;
  const sql = "SELECT * FROM users WHERE username = '" + username + "'";
  const results = dbQuery(sql);
  res.json(results);
});
\end{lstlisting}

Models reported SQL injection in 25 of 25 \texttt{js-project-tigerteam} model runs. In this fixture, Snyk Code was right not to report it: \texttt{dbQuery()} logs the string and returns an empty array. There is no executable SQL sink. This is the kind of case where an LLM can mistake vulnerability-shaped code for an exploitable vulnerability.

\texttt{js-project-nightowl} showed the opposite lesson. All 25 model runs reported SQL injection outside the Snyk Code reference set, and this time the model signal is likely valuable:

\begin{lstlisting}[language=JavaScript]
deleteTodo: (id) => db.prepare("DELETE FROM todos WHERE id = " + id).all(),
\end{lstlisting}

That finding was counted as unmatched because it was not in the Snyk Code reference set. It should not be dismissed as hallucination. It is likely a real product gap to investigate. The benchmark is stronger, not weaker, because it exposed that case.

\begin{figure}[htbp]
\centering
\includegraphics[width=1.0\linewidth]{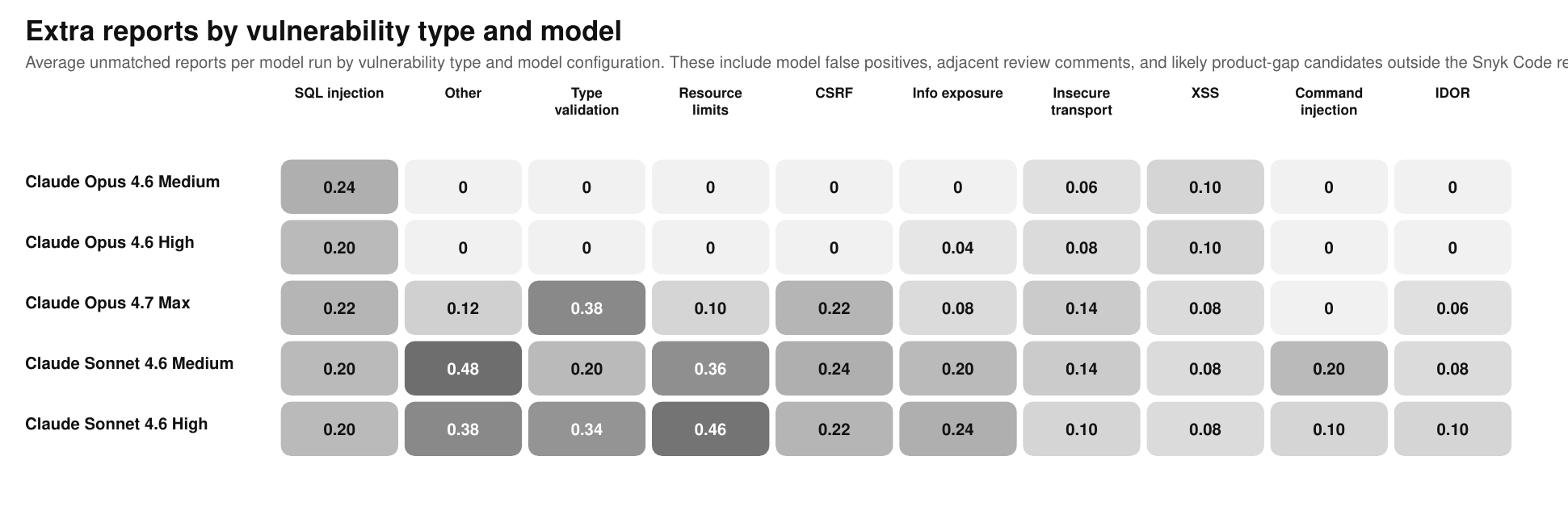}
\caption{Average unmatched reports per model run by vulnerability type and model configuration. These include model false positives, adjacent review comments, and likely product-gap candidates outside the Snyk Code reference set.}
\label{fig:extra-reports}
\end{figure}

This second heatmap shows the other side of complementarity: extra model reports are not one homogeneous category. Some are likely false positives, like the non-executable SQL-shaped mock helper in \texttt{js-project-tigerteam}. Some are adjacent security review comments that are out of scope for the reference set. Some, like the SQL injection report in \texttt{js-project-nightowl}, are likely valid findings that should feed back into Snyk Code coverage.

The complementarity also runs in the other direction. \texttt{js-project-nightowl} is the most app-like fixture in JS 1.0: \texttt{server.js} is 198 lines, with \texttt{db.js} and \texttt{public/app.js} adding another 183 lines of JavaScript. It has routing, uploads, attachment deletion, downloads, and database state. Claude Opus 4.6 High was perfectly stable on this fixture, but stable at only 40.0\% Snyk-reference F1. Across five repetitions it missed every path-traversal reference finding and two of three resource-limit finding opportunities.

\begin{quote}
\textbf{Highlight:} In the largest app-like fixture, Claude Opus 4.6 High was the best model at only 40.0\% Snyk-reference F1, repeatedly missing path traversal and resource-limit vulnerabilities.
\end{quote}

\begin{figure}[htbp]
\centering
\includegraphics[width=0.95\linewidth]{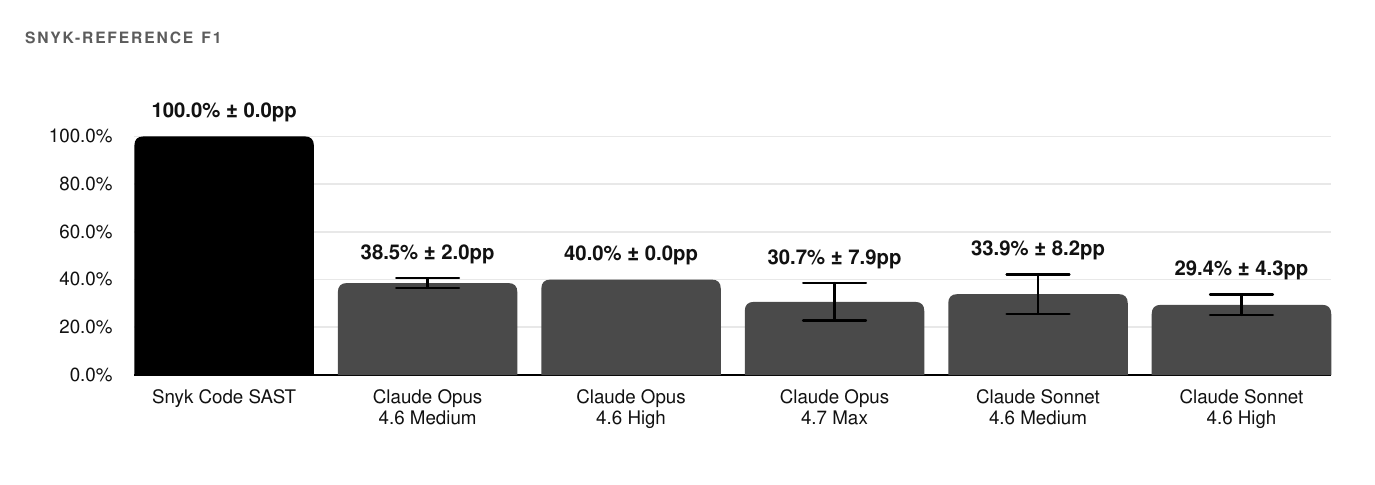}
\caption{Mean benchmark score for the larger multi-file fixture. Error bars show standard deviation across repeated runs.}
\label{fig:larger-fixture}
\end{figure}

The missed pattern spanned repeated attachment flows:

\begin{lstlisting}[language=JavaScript]
if (req.file) {
  if (existing.attachment_stored_name) {
    fs.unlink(path.join(UPLOADS_DIR, existing.attachment_stored_name), () => {});
  }
  updates.push("attachment_original_name = ?", "attachment_stored_name = ?");
  values.push(req.file.originalname, req.file.filename);
} else if (req.body.removeAttachment === true || req.body.removeAttachment === "true") {
  if (existing.attachment_stored_name) {
    fs.unlink(path.join(UPLOADS_DIR, existing.attachment_stored_name), () => {});
  }
}
\end{lstlisting}

The model found some representative issues, then failed to enumerate repeated vulnerable sinks. That is exactly where deterministic data-flow analysis is valuable. SAST coverage and model review are not duplicates of each other; they are different instruments with different blind spots.

\subsection{More Expensive LLM Runs Did Not Mean Better Coverage}

Claude Opus 4.7 Max was the most expensive model configuration in this run, but not the best performing one. It averaged 95,969 tokens and \$0.3559 per model session. Claude Opus 4.6 Medium averaged 51,574 tokens and \$0.0628 per model session. Opus 4.7 Max therefore cost 5.67x more and used 1.86x more tokens, while scoring lower: 68.8\% Snyk-reference F1 versus 75.4\% for Opus 4.6 Medium.

\begin{quote}
\textbf{Highlight:} Claude Opus 4.7 Max cost 5.7x more than Claude Opus 4.6 Medium, used 1.9x more tokens, and scored lower.
\end{quote}

\begin{figure}[htbp]
\centering
\includegraphics[width=0.95\linewidth]{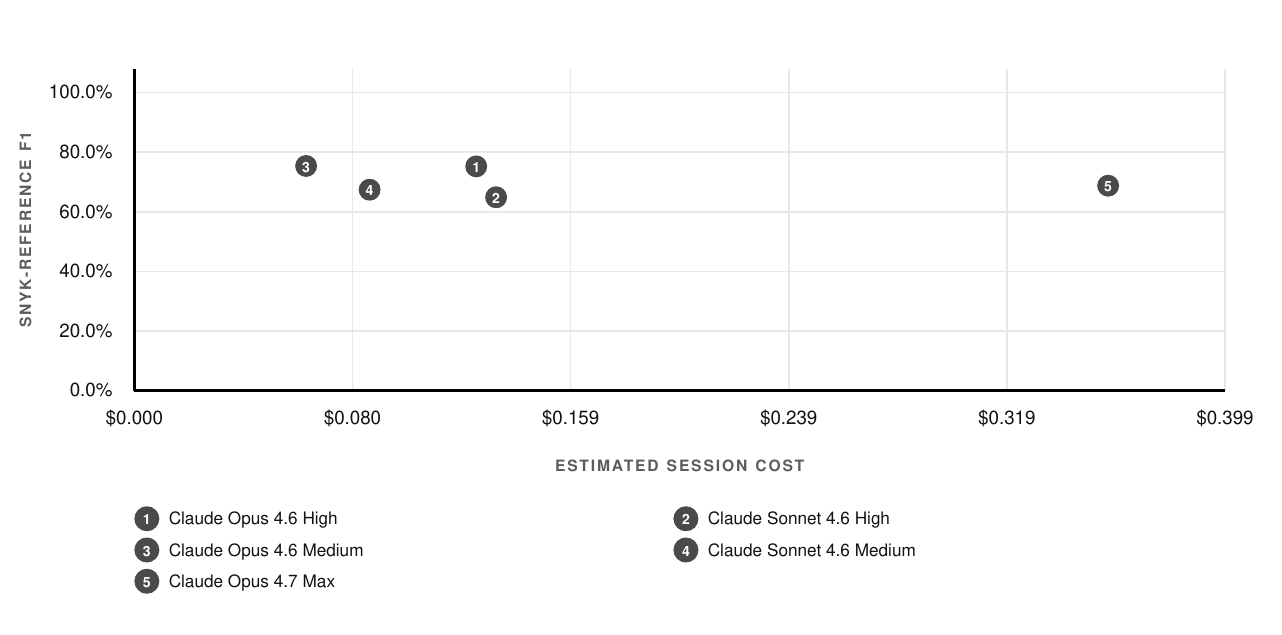}
\caption{Model-only cost/quality tradeoff. Better points move toward the top-left: higher Snyk-reference F1 at lower estimated model-session cost.}
\label{fig:score-vs-cost}
\end{figure}

The absolute dollar amounts are small because the fixtures are small. The scaling question is not. Real security checks run during coding-agent sessions, commits, pull requests, and continuous-integration jobs across repositories that are orders of magnitude larger than these snippets. More expensive inference is not automatically better security coverage.

\subsection{Agreement Scores Against the Snyk Code Reference Set}

Snyk-reference F1 is still useful, as long as it is described precisely: it measures agreement with the Snyk Code reference set. On that metric, Snyk Code SAST reproduced its reference set with 100.0\% Snyk-reference F1 and 0.0 percentage-point score standard deviation. The best model configuration was Claude Opus 4.6 Medium at 75.4\% Snyk-reference F1, 68.0\% recall, and 91.5\% precision.

\begin{table}[htbp]
\centering
\caption{Agreement scores, runtime, token use, and estimated model-session cost.}
\label{tab:agreement}
\small
\resizebox{\textwidth}{!}{%
\begin{tabular}{lrrrrrrr}
\toprule
Configuration & F1 & F1 std. dev. & Recall & Precision & Avg. duration & Avg. tokens & Est. cost \\
\midrule
Snyk Code SAST & 100.0\% & 0.0 pp & 100.0\% & 100.0\% & 14.8s & 0 & N/A \\
Claude Opus 4.6 Medium & 75.4\% & 0.2 pp & 68.0\% & 91.5\% & 27.3s & 51,574 & \$0.0628 \\
Claude Opus 4.6 High & 75.2\% & 0.3 pp & 68.2\% & 89.8\% & 53.8s & 66,929 & \$0.1249 \\
Claude Opus 4.7 Max & 68.8\% & 2.2 pp & 71.4\% & 69.6\% & 37.4s & 95,969 & \$0.3559 \\
Claude Sonnet 4.6 Medium & 67.4\% & 0.9 pp & 80.9\% & 62.6\% & 59.3s & 56,992 & \$0.0860 \\
Claude Sonnet 4.6 High & 64.9\% & 3.5 pp & 81.3\% & 58.6\% & 94.8s & 74,240 & \$0.1322 \\
\bottomrule
\end{tabular}%
}
\end{table}

Table~\ref{tab:agreement} should not be read as ``Snyk proved Snyk is 100\% accurate.'' It should be read as: Snyk Code produced a deterministic reference set; models partially agreed with it; and the differences reveal repeatability, cost, and coverage tradeoffs worth measuring.

\section{Discussion}

The reference set comes from Snyk Code. That is transparent and reproducible, but circular if treated as a universal truth set. This report avoids that claim. The benchmark measures model agreement with Snyk Code findings and uses divergences to study repeatability and complementarity.

The scorer is generous. It matches by vulnerability type, not exact file, line, severity, or source-to-sink identity. A stricter scorer would likely reduce model agreement scores and expose more duplicate-flow mistakes.

The fixtures are small JavaScript and Express applications. They are useful for controlled measurement, but they do not cover large monorepos, framework-heavy TypeScript applications, multi-service architectures, or business-logic vulnerabilities. \texttt{js-project-nightowl} already shows that app-like structure changes model behavior.

The recurrence analysis uses normalized finding signatures. For the model-by-model unmatched charts, the signature is task + vulnerability type + file + line, grouped separately for each model configuration. Different normalization choices change the exact percentages, which is why the signature is part of the chart handoff and reproducibility notes.

\section{Conclusion}

Snyk VulnBench JS 1.0 shows that agentic LLM security review can be useful, but its output is not uniformly repeatable. When models found Snyk Code reference findings, those findings were usually stable across repeated runs. When models produced findings outside the reference set, the review queue was much noisier: nearly half of unique unmatched reports appeared in only one of five identical scans.

The next Snyk VulnBench release should move beyond small self-contained snippets. We plan to add more full-fledged application structures, LLM-sourced vulnerabilities, business-logic and broken-object-level-authorization classes, and an independent ground truth source such as BaxBench-style reference data \cite{vero2025baxbench}.

Future reports should also evaluate combined workflows: model-only review, SAST-only analysis, and LLM review augmented with SAST context. The JS 1.0 data already points in that direction. Models and SAST do not fail the same way. That is the reason to combine them.

\bibliographystyle{plain}
\bibliography{references}

\end{document}